\documentclass{acm_proc_article-sp}
\usepackage {verbatim}
\usepackage {graphicx}
\usepackage {multirow}
\usepackage {rotating}

\begin{document}

\title {Distributed Multi Class SVM for Large Data Sets}

\permission {Permission to make digital or hard copies of all or part of this work for personal or classroom use is granted without fee provided that copies are not made or distributed for profit or commercial advantage and that copies bear this notice and the full citation on the first page. Copyrights for components of this work owned by others than ACM must be honored. Abstracting with credit is permitted. To copy otherwise, or republish, to post on servers or to redistribute to lists, requires prior specific permission and \/or a fee. Request permissions from Permissions@acm.org. \\
WCI `15, August 10 - 13, 2015, Kochi, India \\
\copyright 
2015 ACM. ISBN 978-1-4503-3361-0/15/08\$15.00 \\
DOI: {http://dx.doi.org/10.1145/2791405.2791534
}}

%
%
%
%
%

\numberofauthors{3} 
%
\author{
%
%
\alignauthor
Aruna Govada\\
       \affaddr{BITS-Pilani, KK Birla Goa Campus}\\
       \affaddr{Goa}\\
       \affaddr{India}\\
       \email{garuna@goa.bits-pilani.ac.in}
\alignauthor
Bhavul Gauri \\
       \affaddr{BITS-Pilani, KK Birla Goa Campus}\\
       \affaddr{Goa}\\
       \affaddr{India}\\
       \email{bhavul93@gmail.com}
\alignauthor
S.K.Sahay\\
       \affaddr{BITS-Pilani, KK Birla Goa Campus}\\
       \affaddr{Goa}\\
       \affaddr{India}\\
       \email{ssahay@goa.bits-pilani.ac.in}
}


\maketitle
\begin{abstract}
Data mining algorithms are originally designed by assuming the data is available at one centralized site.These algorithms also assume that the whole data is fit into main memory while running the algorithm. But in today's scenario the data has to be handled is distributed even geographically. Bringing the data into a centralized site is a bottleneck in terms of the bandwidth when compared with the size of the data. In this paper for multiclass SVM  we propose an algorithm which builds a global SVM model by merging the local SVMs using a distributed approach(DSVM). And the global SVM will be communicated to each site and made it available for further classification. The experimental analysis has shown promising results with better accuracy when compared with both the centralized and ensemble method. The time complexity is also reduced drastically because of the parallel construction of local SVMs. The experiments are conducted by considering the data sets of size 100s to hundred of 100s which also addresses the issue of scalability.

\end{abstract}




\section{Introduction}
\par Data mining algorithms demand the data to be available at one centralized site. In today's era of massive data sets which are distributed geographically, bringing this whole data to a centralized site is almost impossible due to the limited bandwidth when compared with the size of the data. And also solving a large problem at a central site is not feasible in terms of the computational complexity. 
\newpage
All traditional data mining algorithms assume that the data should fit into main memory which is a challenge for data mining algorithms in terms of scalability. \cite{tan:vipin} 
\par In many domains like financial\cite{hong:wang}, health care\cite{chen:young}, astronomy\cite{sar:mal} the data is overflowing resulting in data avalanche due to the advances in data collection methodologies. In all these applications, the data may not reside at a centralized location. For example, the different sky survey telescopes which are geographically distributed must be collecting the data of common interest continuously. Mining from these massive data can not be achieved unless the data mining algorithms are capable of handling the decentralized data.\cite{das:bha} \cite{hil:chr}.
\par Distributed Data Mining (DDM) can be one of the solution for the above said problem. DDM can be achieved in two ways, Data Mining on distributed data or Distributing the Data Mining on the centralized data. In this paper we discuss the first scenario in which the data is distributed at different sites. DDM is possible on horizontal partition as well as vertical partition also. In Horizontal partition the number of attributes are constant at all n different locations but the number of instances may vary. Whereas in vertical partition the number of instances are constant at all n different locations but number of attributes may vary. \cite{dut:h}
\par In this paper the data is partitioned in horizontal manner. The proposed distributed approach is compared with the centralized method by bringing the distributed data to one central site. The multiclass SVM is achieved using One-Versus-One(OVO) approach both in centralized and distributed approach. The experimental analysis shows how our distributed approach is better than the normal approach in terms of accuracy,training time and testing time.  Experiments are conducted by considering different data sets of different size. Our proposed approach could succeed in building the global SVM in case of large data sets whereas the centralized approach could not handle the data to build the training model.
\par The rest of the paper is organized as follows. In next section the related work is discussed. Section 3 briefly describe the binary SVM, OVO multiclass SVM. In section 4 we present our distributed approach for scalable distributed data mining to construct the merged model. In section 5 we present the experimental analysis of our algorithm. At the end in section 6 the conclusions of the paper are mentioned.

\section{Related Work}

Though a decent amount of work has been done in multiclass SVM and parallel/distributed binary SVM individually, the research of distributed multiclass SVM needs more exploration by the research community. There is a continuous attention on SVM because it was proved as the best method in several applications even though it is computationally expensive \cite{ren:jas}.

In 1998 Han.X et al. discussed the model of coupling the estimation of class probabilities for each pair of classes \cite{han:berg}. The used classifiers include linear discriminant, nearest neighbors and SVM.

In 2005 Hian et al. discussed the association between the symptoms and the treatment of a patient \cite{hian:tan}. their discussion also includes the significance of handling the huge amount with the help of data mining.

In 2010 Stefano et al. discussed the construction of a SVMs based on the Minimal Enclosing Ball(MEB) when the training data is partitioned at several locations \cite{ste:nan}. It is shown how the union of local core-sets provides a close approximation to a global core-set from which the SVM can be recovered.

In 2011 Ahmed et al. designed a hybrid ensemble model for credit risk which combines both clustering and classification \cite{ahmd:gho}. In this SVM classifiers are the members in the ensemble model. 

A multiclass classification approach for large data sets is discussed by using SVM,enclosing ball(MEB) method \cite{jair:yu}.
Solving a single optimization problem for the multiclasses is very expensive in terms of the time. A distributed parallel training approach is discussed for single-machine problem in \cite{hanx:berga}.

In 2014 Aruna.G et al. proposed a binary tree based support vector machine \cite{aru:ran} which reduces the number of binary classifiers when compared with OVO and OVA approaches. This algorithm is implemented in a distributed manner under HADOOP framework.

\section{Preliminaries}
\subsection {Binary SVM}
\par Given some training data D, a set of k points of the form
 D = {\{($p_i$,$q_i$) | $p_i \in R^p ,q_i \in \{-1,1\}$\}} , i=1 to k \\ 
The goal is to find the maximum-margin hyperplane that divides the points having $q_i=1$ from those having $ q_i=-1$. \\

\par The function of learning  in binary SVM can be represented as follows. \cite{tan:vipin}

$$min_w = \frac{\parallel w \parallel^2}{2}$$   
subject to $q_i(w.p_i + b) \ge 1, \quad i =1,2,....k$  where $w$ and $b$ are the parameters of the model for total $k$ number of instances.\\

 Using Lagrange multiplier method the following equation has to be solved,

              $$ L_p = \frac{\parallel w \parallel^2}{2} - \sum_{i=1...k} \lambda_i (q_i(w.p_i   + b) -1)$$ \\ 
            The dual version of the above problem is \\
    $$ L_D = \sum_{i=1...k} \lambda_i - \frac{1}{2} \sum_{i,j} \lambda_i \lambda_j q_iq_jp_i.p_j $$ 
    subject to $$\lambda_i \ge 0 $$
     $$ \lambda_i (q_i(w.p_i   + b) -1)=0 $$ 
where $\lambda_i $ are known as the Lagrange multipliers.\\
By solving this dual problem, SVM(hyperplane) will be found. Once the training model is built,the class label of a testing object z can be predicted as follows.\\
   $$  f(z) = sign\sum_{i=1...k} (\lambda_i q_i p_i.z + b) $$ 
   if $ f(z) \ge 0 $ z will be predicted as +ve class else -ve class.

\subsection{One-Versus-One(OVO):A MultiClass SVM}
\par Multiclass classification is the problem of classifying instances into one class label among the N-class labels.
Build N(N-1)/2 classifiers one classifier to distinguish each pair of classes i and j. Let $f_{ij}$ be the classifier where class i were +ve examples and class i were -ve. Classify using 

   $$ f(x) = \underset{i}{\arg\max} \Bigg( \sum _j f_{ij} (x) \Bigg) $$

\par This way in OVO, each class is compared to every other class. A binary classifier is built to differentiate between each pair of classes, while discarding the rest of the classes. When an unseen object has to be classified into one of the class, a voting is made among the classifiers and the class with the maximum number of votes will be considered as the best choice.
   
\section{The proposed Approach} 

\par The data be distributed among  n sites with equal number of attributes but varied in number of instances.  
\begin {enumerate} 

\item Let the data is distributed among n sites .

 $$[\mathbf{X}]_{p \times q} = (X_1, X_2, X_3, ......... X_n)$$
                 
where data $X_j$ is a ${p_j \times q}$ matrix residing at the site $S_j$ and $p = \sum_{j = 1}^{n} p_j$

\item Build the local SVM models $SVM_1,SVM_2....SVM_n$ at all n sites individually.

\item Construct a directed graph as follows.
\begin {itemize}
\item Each site is a vertex .
\item Edge (i->j)  refers to the training model $SVM_i$  w.r.t test data at site j .( where  i= 1 to n , j=1 to n , $i \ne j$ )
\item Label the edge with Accuracy of ( $SVM_i$ ,j) .
\end {itemize}

// Merging the local models into a global model

\item For each vertex j, Find out the Maximum labeled edge among all the edges from i to j, where  i= 1 to n , $i \ne j$. and store the values in ${n \times 2}$ matrix as follows.
\begin{itemize}
     \item For (k= 1 to n) 
 \begin{itemize}    
 \item Best [k][1]=i;
\item Best [k][2]=j;  // Decides the best model $SVM_i$   w.r.t  test data at site j
\end {itemize}		 
\end {itemize}
\item Find out the element  which is having the maximum frequency among Best[i][1] , where i=1 to n. And the corresponding SVM  model is decided as the global/merged model.
\end {enumerate}

\par The architecture of the proposed approach is shown in figure 1, where the global model is built by merging the local SVMs. And the global model made available at each local site by transmitting it so that it can be used in future for classifying unseen objects.
\subsection{Graphical Representation}

\par The training data at n sites can be constructed as a graph and shown in Figure 2. Each site is considered as a vertex and the edge from the vertex i to vertex j represents the accuracy of the SVM model of the training data at site i w.r.t the test data at site j .\\

The \emph{Accuracy Matrix} $A_{ij}$ is an 
$n \times n$~matrix which can be written as follows.
\[ \left( \begin{array}{cccccc}

a_{11} & a_{12} & a_{13} & {..} & {..} & a_{1n}\\
a_{21} & a_{22} & a_{23} & {..} & {..} & a_{2n}\\
a_{31} & a_{32} & a_{33} & {..} & {..} & a_{3n}\\
{..}& {..} & {..} & {..} & {..} & {..} \\
{..}& {..} & {..} & {..} & {..} & {..} \\
a_{n1} & a_{n2} & a_{n3} & {..} & {..} & a_{nn}
 \end{array} \right)\] 
To get the final global SVM find out the Maximum of each column and note down the corresponding i value of SVM. Among these n-Max values (n-SVMs), choose the SVM which will get the maximum count as the final global model.

\begin{figure}\centering
\includegraphics[scale=0.5]{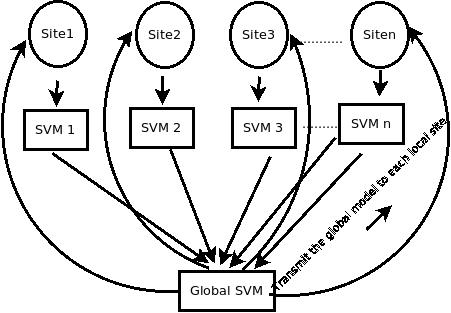}
\caption {The Architecture}
\label{fig:ex}
\end{figure}

\begin{figure}\centering
\includegraphics[scale=0.5]{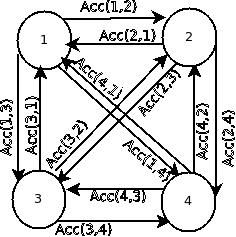}
\caption{SVMs of $i^{th}$ site w.r.t test data of $j^{th}$ site.}
\label{fig:ex}
\end{figure}

\section {Experimental Section}

\par We implemented the algorithm on three data sets Mfeat-Fac, Pendigits and  Sloan Digital Sky Survey (SDSS). Mfeat and Pendigit are taken from UCI machine learning repository https://archive.ics.uci.edu/ml/datasets.html. SDSS is an astronomical catalog taken from http://skyserver.sdss.org /dr7/en/tools/search/sql.asp based on different conditions of its attributes. Mfeat, Pendigit and SDSS data sets are divided into 3,4,4 partitions respectively and implemented DSVM.
\par In our analysis the DSVM is compared with centralized SVM and Ensemble SVM. The results show the better accuracy with reduced training time and testing time. 

\par In table 1 the description of the data sets that are considered for analysis is given. In table 2 the training accuracy and the training time is given. The error of global model will not exceed the error of Max(Local Models) as the data is independent where as in ensemble model this is not guaranteed because the same samples can be repeated at different sites. 
In table 3, 4, 5 the construction of the global models of Mfeat-Fac, Pendigits, SDSS data sets are given respectively. 
If we observe ,
\par The accuracy matrix of Mfeat-Fac with three sites
\[ \left( \begin{array}{ccc}
-1 & 96 & 96.8 \\
98.2 & -1 & 98 \\
97 & 96.3 & -1  \\
\end{array} \right)\] 
The best model will be $SVM_2$ as it has the maximum count and is shown in table 3.
 
 \par The accuracy matrix of Pendigits with four sites 
\[ \left( \begin{array}{cccc}
-1 & 99.45 & 99.70 & 99.30 \\
99.50 & -1 & 99.60 & 99.40 \\
99.60 & 99.25 & -1 & 99.20 \\
99.50 & 99.35 & 99.5 & -1 \\
 \end{array} \right)\] 
The best model will be $SVM_1$ as it has the maximum count and is shown in table 4.
 
\par The accuracy matrix of SDSS with four sites 
\[ \left( \begin{array}{cccc}
-1 & 68.38 & 68.39 & 68.52 \\
37.45 & -1 & 89.16 & 89.02 \\
37.45 & 89.53 & -1 & 89.37 \\
37.47 & 89.63 & 89.66 & -1 \\
 \end{array} \right)\] 
The best model will be $SVM_4$ as it has the maximum count and is shown in table 5.
\par In table 6 the ensemble method is computed for all 3 data sets. In ensemble method the test data has to be tested every time with all the available training models and the class label will be decided by the voting approach. Where as in DSVM , the final global model is merged from all local models. And whenever an unseen object has to be classified, it will be tested with only one global model. Hence the testing time of DSVM is reduced when compared with Ensemble SVM.

Finally in Table 7, DSVM is compared with  OVO multiclass SVM with centralized and ensemble model. The results show that the accuracy of DSVM is equivalent to centralized SVM with reduced training time and testing time. The training time of DSVM is considered as the time of Max(local SVMs) as the local SVMs can be constructed in a parallel manner. For the data set SDSS the system crashed during the training of SVM for centralized method but DSVM built the training model without any hurdle, hence it is scalable.

\begin{table*}
\caption{The description of the data sets used.}
\centering

\begin{tabular}{|c|c|c|c|c|c|c|c|c|}
\hline
Dataset & Features & Training Sizes&Testing Size & Class Labels & At site1 & At site 2 & At site 3 & At site 4 \\
\hline
Mfeat-Fac & 216 & 1800 & 200 & 10 & 500 & 800 & 500 & - \\
\hline
Pendigits & 16 & 7514 &3478 & 9 & 1800 & 2000 & 1494 & 2200 \\
\hline
SDSS & 5 & 175000& 1000 & 5 & 20000 & 30000 & 50000 & 75000 \\
\hline
\end{tabular}

\end{table*}

\begin{table*}
\caption{Training Accuracy and Training Time of Data sets at Distributed Sites.}
\centering
\begin{tabular}{|c|c|c|c|c|c|c|c|c|}
\hline
 Dataset &\multicolumn{2}{c|}{Site 1}&\multicolumn{2}{c|}{Site 2}&\multicolumn{2}{c|}{Site 3} & \multicolumn{2}{c|}{Site 4}\\
\cline{2-9}
&Accuracy&Time&Accuracy&Time&Accuracy&Time&Accuracy&Time\\
\hline
Mfeat-Fac & 100 & 0.1365 & 100 & 0.256 & 100 & 0.143 & - & - \\
\hline
Pendigits & 100 & 0.062 & 99.53 & 0.110 & 100 & 0.049 & 100 & 1.0132 \\
\hline
SDSS & 100 & 49.996 & 89.09 & 115.76 & 89.13 & 369.93 & 89.76 & 1728.097 \\
\hline
\end{tabular}
\end{table*}

\begin{table}
\caption{The Best Global Model of Mfeat-Fac:${SVM_2}$ }
\centering
\begin{tabular}{|c|c|c|c|}
\hline
Test Data & Training Model& Accuracy & Best Model\\
\hline
\multirow{2}{*}{Site$_1$} & $SVM_2$ & 98.2& \multirow{2}*{SVM$_2$} \\
\cline{2-3}
 & $SVM_3$ & 97.0& \multirow{2}*{} \\
\hline
\multirow{2}{*}{Site$_2$} & $SVM_1$ & 96.0& \multirow{2}*{SVM$_3$} \\
\cline{2-3}
 & $SVM_3$ & 96.3& \multirow{2}*{} \\
 \hline
 \multirow{2}{*}{Site$_3$} & $SVM_1$ & 96.8& \multirow{2}*{SVM$_2$} \\
\cline{2-3}
 & $SVM_2$ & 98.0& \multirow{2}*{} \\
 \hline
\end{tabular}
\end{table}

\begin{table}
\caption{The Best Global Model of Pendigits:${SVM_1}$ }
\centering
\begin{tabular}{|c|c|c|c|}
\hline
Test Data & Training Model& Accuracy & Best Model\\
\hline
\multirow{3}{*}{Site$_1$} & $SVM_2$ & 99.50& \multirow{3}*{SVM$_3$} \\
\cline{2-3}
 & $SVM_3$ & 99.60& \multirow{3}*{} \\
 \cline{2-3}
 & $SVM_4$ & 99.50& \multirow{3}*{} \\
\hline
\multirow{3}{*}{Site$_2$} & $SVM_1$ & 99.45& \multirow{3}*{SVM$_1$} \\
\cline{2-3}
 & $SVM_3$ & 99.25& \multirow{3}*{} \\
 \cline{2-3}
 & $SVM_4$ & 99.35& \multirow{3}*{} \\
 \hline
\multirow{3}{*}{Site$_3$} & $SVM_1$ & 99.70& \multirow{3}*{SVM$_1$} \\
\cline{2-3}
 & $SVM_2$ & 99.60& \multirow{3}*{} \\
 \cline{2-3}
 & $SVM_4$ & 99.50& \multirow{3}*{} \\
 \hline
\multirow{3}{*}{Site$_4$} & $SVM_1$ & 99.30& \multirow{3}*{SVM$_2$} \\
\cline{2-3}
 & $SVM_2$ & 99.40& \multirow{3}*{} \\
 \cline{2-3}
 & $SVM_3$ & 99.20& \multirow{3}*{} \\
 \hline
 
\end{tabular}
\end{table}

\begin{table}
\caption{The Best Global Model of SDSS : ${SVM_4}$ }
\centering
\begin{tabular}{|c|c|c|c|}
\hline
Test Data & Training Model& Accuracy & Best Model\\
\hline
\multirow{3}{*}{Site$_1$} & $SVM_2$ & 37.45& \multirow{3}*{SVM$_4$} \\
\cline{2-3}
 & $SVM_3$ & 37.45& \multirow{3}*{} \\
 \cline{2-3}
 & $SVM_4$ & 37.47& \multirow{3}*{} \\
\hline
\multirow{3}{*}{Site$_2$} & $SVM_1$ & 68.38& \multirow{3}*{SVM$_4$} \\
\cline{2-3}
 & $SVM_3$ & 89.53& \multirow{3}*{} \\
 \cline{2-3}
 & $SVM_4$ & 89.63& \multirow{3}*{} \\
 \hline
\multirow{3}{*}{Site$_3$} & $SVM_1$ & 68.39& \multirow{3}*{SVM$_4$} \\
\cline{2-3}
 & $SVM_2$ & 89.16& \multirow{3}*{} \\
 \cline{2-3}
 & $SVM_4$ & 89.66& \multirow{3}*{} \\
 \hline
\multirow{3}{*}{Site$_4$} & $SVM_1$ & 68.52& \multirow{3}*{SVM$_3$} \\
\cline{2-3}
 & $SVM_2$ & 89.02& \multirow{3}*{} \\
 \cline{2-3}
 & $SVM_3$ & 89.37& \multirow{3}*{} \\
 \hline
\end{tabular}
\end{table}

\begin{table}
\caption{The Ensemble Model of Data Sets }
\centering
\begin{tabular}{|c|c|c|c|}
\hline
Data Set & Local Site & Accuracy & Voting Model\\
\hline
\multirow{3}{*}{Mfeat-Fac} & {Site$_1$} & 97.50 & \multirow{3}*{SVM$_2$} \\
\cline{2-3}
 & {Site$_2$} & 98.00 & \multirow{3}*{} \\
 \cline{2-3}
 & {Site$_3$} & 97.00 & \multirow{3}*{} \\
\hline
\multirow{3}{*} {Pendigits} & {Site$_1$} & 10.40 & \multirow{3}*{SVM$_1$} \\
\cline{2-3}
 & {Site$_2$} & 10.37 & \multirow{3}*{} \\
 \cline{2-3}
 & {Site$_3$} & 10.37 & \multirow{3}*{} \\
 \cline{2-3}
 & {Site$_4$} & 10.37 & \multirow{3}*{} \\
\hline
\multirow{3}{*} {SDSS} & {Site$_1$} & 85.80 & \multirow{3}*{SVM$_4$} \\
\cline{2-3}
 & {Site$_2$} & 53.30 & \multirow{3}*{} \\
 \cline{2-3}
 & {Site$_3$} & 53.30 & \multirow{3}*{} \\
 \cline{2-3}
 & {Site$_4$} & 87.10 & \multirow{3}*{} \\
 \hline
\end{tabular}
\end{table}

\begin{table*} [!htbp]
\caption{The Comparison Of the DSVM  with Centralized and Ensemble Models }
\centering
\begin{tabular}{|c|c|c|c|c|c|c|c|c|c|}
\hline
 Dataset & \multicolumn{3}{c|}{Centralized} &\multicolumn{3}{c|}{Ensemble} & \multicolumn{3}{c|}{DSVM} \\
\cline{2-10}
&Accuracy&Tr.Time&Te.Time &Accuracy&Tr.Time&Te.Time &Accuracy&Tr.Time&Te.Time \\
\hline
Mfeat-Fac & 99 & 0.667 & 0.133& 98 & 0.234 & 0.204 & 98 & 0.234 & 0.077  \\
\hline
Pendigits & 10.40 & 0.477 & 0.190 & 10.40 & 0.243 &0.868  &10.40 & 0.243 & 0.135  \\
\hline
SDSS & - & - & - & 89 &227.24&5.52 & 89 & 227.24 & 2.94  \\
\hline
\end{tabular}
\end{table*}

\section{Conclusions}

We propose an algorithm DSVM which builds a global SVM by merging the local SVMs that are distributed at different sites.  Experimental results show that the performance of DSVM is better than the centralized and Ensemble model both in accuracy and training,testing time. DSVM is also capable of handling large data sets, hence scalable. Though Ensemble method also can handle large data sets at the time of training, testing time will be very costly as it has to be tested with every training  model (which are available at different locations) and follow voting mechanism. But DSVM will have only one global model and it is scalable for training as well as for testing also. Further enhancement can be done by considering the vertical partition of the data at different sites.


\section{Acknowledgments}
We are thankful for the support provided by the Department of Computer Science and Informations Systems, BITS, Pilani, K.K. Birla Goa Campus to carry out the experimental analysis.

%
\bibliographystyle{abbrv}
\bibliography{wci}  
\end{document}